\documentstyle[12pt]{article}

\begin{document}

\title{MAGNETIC RECONNECTION: SWEET-PARKER VERSUS PETSCHEK}
\author{RUSSELL M. KULSRUD\\
{\it Princeton Plasma Physics Laboratory}\\
rmk@pppl.gov}

\maketitle

\abstract

The two theories for magnetic reconnection, one
of Sweet and Parker, and the other of Petschek,
are reconciled by exhibiting an extra condition
in that of Petschek which reduces his theory to
that of Sweet and Parker, provided that the
resistivity is constant in space. On the other
hand, if the resistivity is enhanced by instabilities,
then the reconnection 
rate of both theories is increased
substantially, but Petschek's rate can be
faster.  A different formula from the usual one
is presented for enhanced Petschek reconnection.

\section{Introduction}

	As is well known, the process of magnetic
reconnection is important in many space and
astrophysical contexts.  The initial problem that
first inspired research into the subject was the
solar flare phenomenon, in which it appeared
that energy was first slowly built up and
stored in the magnetic field, and then suddenly 
released into thermal 
and kinetic energy.  The first solution of
the problem was given independently by Sweet 
(1958) and
Parker (1957), who approximated the problem as
a two dimensional incompressible MHD problem.
They showed that the problem was
essentially a boundary layer problem, and they
estimated the rate of reconnection from
a boundary layer analysis. This boundary layer
analysis led to release of
magnetic energy over a period of time several
orders of magnitude longer  than the observed
energy release time in solar flares.  A probable
explanation of this discrepancy could be the
fact that their estimates of the reconnection
rate are based on  normal (Spitzer) resistivity,
while in the actual 
solar flare the resistivity could be greatly enhanced,
leading to a much faster energy release.

	On the other hand, at the time when
Sweet and Parker developed their theories, the possibility
of enhanced resistivity
was not appreciated, and other means of increasing
the reconnection rates were sought.  Petschek (1963)
pointed out that, since the magnetic reconnection
was a topological process, the field lines
need not reconnect resistively along the
entire length of the boundary layer, but
could merge over a shorter length $ L' $.

	 For this to happen, the rest of the
boundary layer region should consist of slow shocks
that could accelerate the matter that did not pass
through the diffusive region.  He found that
the resulting reconnection rate was increased by 
the factor $ \sqrt{L/L'} $, and by choosing
$L' $ small enough, a very rapid reconnection could
be achieved.  

	Ever since the two theories, Sweet-Parker's
and Petschek's, were published, there has been a
controversy over which one
was the correct one to apply.  The controversy
seemed to be settled, by the
rather complete numerical simulation
of Biskamp (1986),  to be in favor of the Sweet-Parker result.
Since then, a number of numerical simulations have 
confirmed this.

	Since both theories seemed rather well founded,
it is a question of how either of them could be
incorrect.  In this note, I  will show that the Petschek
theory, as he proposed, it is indeed not correct,
at least in the context of MHD with constant resistivity.
In the development of his theory, Petschek left out
one condition that also must be satisfied.
The satisfaction of this condition leads to 
a unique determination of the length $ L' $ 
in his theory, and indeed, if the resistivity is
constant in space,
 it is the
case that $ L' $ is equal to $L$. 
This reduces Petschek's enhancement factor
$\sqrt{L/L'} $ to unity and Petschek's 
reconnection rate to Sweet-Parker's
rate for constant resistivity.

	On the other hand, if one considers
the possibility of enhanced resistivity,
two things happen.
(1) The Sweet-Parker
 reconnection rate becomes much faster,
for the solar flare case, and (2) because
such enhanced  resistivity is very sensitive to current
density, it can be space dependent also.  This leads to 
$L' $, being  smaller than $L$, and to a 
even faster, Petschek like, reconnection rate.

\section{The Boundary Layer}

\begin{figure}[t]
\vspace{11.6cm}
\includegraphics{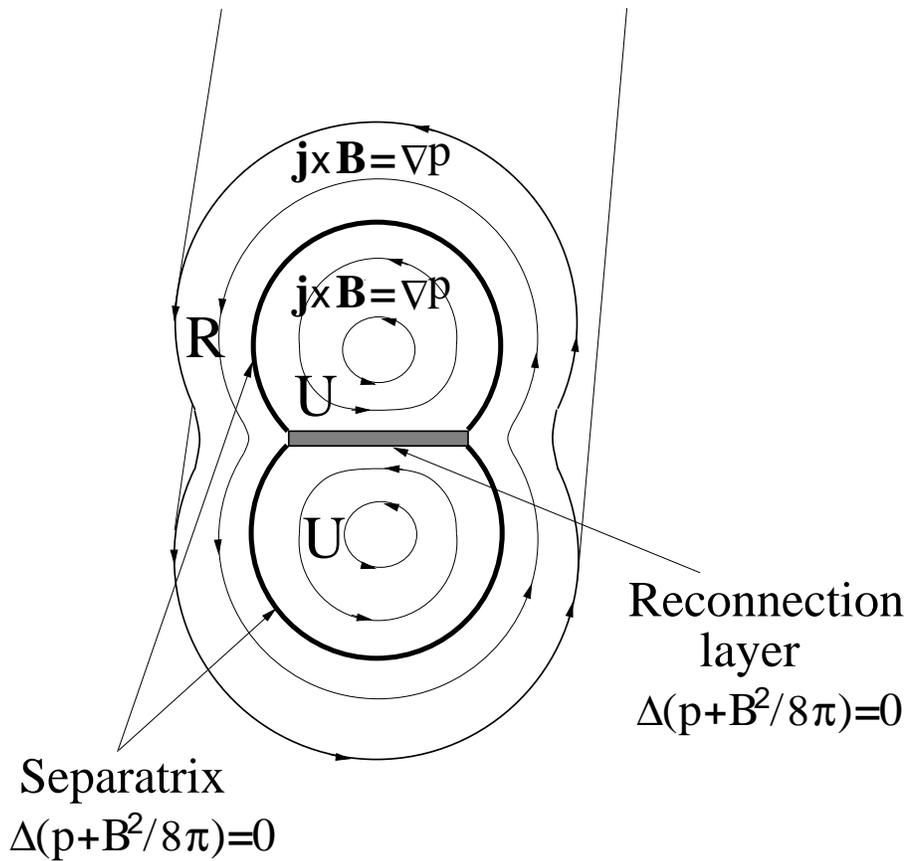}
\caption{Slow merging of two cylinders.
The situation is in a slowly evolving
quasistatic equilibrium everywhere,
except in the layers across which
$ p + B^2/8 \pi $ is continuous.
$ R $ is the region of reconnected
flux, and $ U $ is the region of
unreconnected flux.  }
\end{figure}

	Reconnection is as much a global phenomena
as a local one.  For example, consider the magnetic
reconnection of two cylinders with opposite 
poloidal flux. (Figure 1.) Let us also assume that
the velocities induced by magnetic reconnection
are slow compared to the Alfven speed everywhere,
except in the reconnection and separatrix 
layers.  Then,   everywhere else, we have
\begin{equation}
\bf{ j \times B } = \nabla p
\end{equation}
Also, since the layers are thin, we have the jump in 
$p + B^2/8 \pi $ zero across these layers.
This means that, if the amount of reconnected
and unreconnected flux is given, and
if the rotational
transform  and pressure are known
on each magnetic surface
as functions of the poloidal magnetic flux,
then there is a unique equilibrium.
But as magnetic reconnection proceeds, one
can keep track of the pressure and
the rotational transform in both the
regions of reconnected and
unreconnected flux.  The two  regions
change geometrically and 
 physically as reconnection proceeds,
Thus, the plasma first moves from the 
unreconnected region into the 
reconnection layer, where it is heated,
then it flows   into the separatrix region,
and, finally, as the magnetic configuration
changes, into the reconnected region.
(However, this does not change the uniqueness
of the equilibrium at each stage of reconnection.)

Because of this uniqueness, the length of the 
reconnection layer $ L $ is totally determined
at each stage as well as the horizontal field
just outside of the layer.

	Appreciating this fact, all three authors
took the length of the layer as well as the
variation of the bounding field as given.
They assumed that the region into which the 
plasma flowed (the separatrix region) was
at the same pressure as the upstream ambient
pressure.  This was the boundary layer
problem to be solved.

\section{The Sweet-Parker and Petschek Theories
for a {Constant} Resistivity}

\begin{figure}[t]
\vspace{6.6cm}
\includegraphics{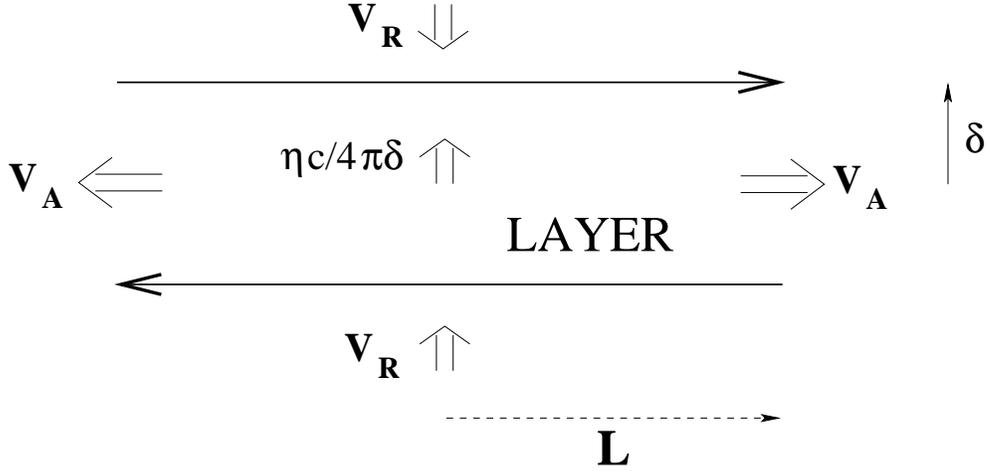}
\caption{The Sweet-Parker layer.  In a
steady state the magnetic diffusion
velocity $ \eta c/4 \pi \delta $ 
balances the incoming reconnection
velocity $ V_R $, and the inflowing mass
$ 4 V_R L $ balances the outgoing
mass $ 4 \delta V_A $. }
\end{figure}

	Let us now examine the two theories.
First consider that of Sweet and Parker.
The reconnection
layer is sketched in figure 2.  It is easily
shown that the flow out of the layer 
is at the Alfven speed, $ V_x = V_A \equiv
B_0/\sqrt{4 \pi \rho } $.  The incoming 
flow of matter $ -L V_y = L V_R $ 
must balance the outgoing flow $V_A \delta$,
where $V_R $, the reconnection velocity,
 is the incoming
velocity outside  of the layer, where
the plasma is tied to the field lines.
$ \delta $ is the half thickness of the layer.
Thus,
\begin{equation}
V_R L = V_A \delta  .
\end{equation}

	On the other hand, by Ohm's law the 
field diffuses up stream with a velocity 
$ \eta c/4 \pi \delta $ with respect to the 
incoming plasma, with velocity $ - V_R $.
Thus, in steady state,
\begin{equation}
V_R = \eta c / 4 \pi \delta.
\end{equation}
From these equations we obtain
\begin{equation}
V_R = \sqrt{\frac{V_A \eta c}{ 4 \pi L}} = \frac{V_A}{S}  ,
\end{equation}
and
\begin{equation}
\delta= \frac{L}{\sqrt{S}}   ,
\end{equation}
where
\begin{equation}
S = \frac{L V_A }{ \eta c/ 4 \pi}   .
\end{equation}
This is the Sweet-Parker result.

\begin{figure}[t]
\vspace{9.5cm}
\includegraphics{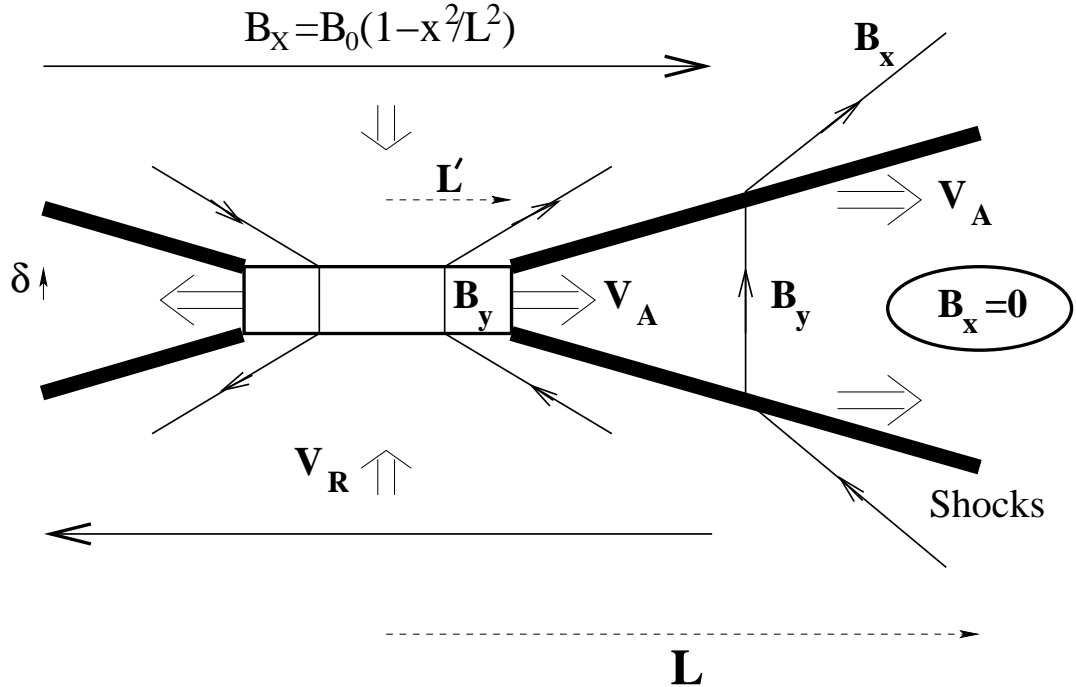}
\caption{The Petschek model.  The diffusion
region of length $ L' < L $ is the 
same as figure 2, with $ L $ taken
equal to $ L' $. Outside the diffusive
region \, shocks accelerate the plasma 
and reduce $ v_x $ to zero.  The shocks'
$ y $ velocity $ B_y/\sqrt{ 4 \pi \rho} $
balances the incoming $ V_R $ velocity
in a steady state.  The $ B_y $ field 
is vital to this model. }
\end{figure}

	The Petschek theory is indicated in
figure 3.  In this model the diffusive
region, in which the merging actually
takes place is of a 
much shorter,  length $L' $,
than $L$. The remaining
length of the boundary is occupied by slow
shocks.  In the diffusive layer
the behavior is similar to the Sweet-Parker
layer, the main difference being that the 
acceleration of the velocity
up to the Alfven speed along the layer,
 is accomplished by 
magnetic tension associated with 
a transverse field component $B_y $.
(In the Sweet-Parker theory this acceleration
is produced mainly by a pressure gradient.)
Outside of the Petschek diffusive layer
the acceleration up to $ V_A $ is accomplished
almost instantaneously by the slow shocks.
The Sweet-Parker model for their diffusive
layer is replaced by the identical conditions
for the Petschek model, 
but with $L $ replaced by $ L' $, leading
to the Petschek reconnection velocity,
\begin{equation}
V_R = \sqrt{\frac{V_A \eta c}{4 \pi L'}} =
\frac{V_A}{\sqrt{S}} \sqrt{\frac{L}{L'}},
\end{equation}
a factor of $\sqrt{L/L'} $ faster than the 
Sweet-Parker reconnection velocity.
The shocks in the outer $ L-L' $ region
reduce the upstream $B_x $ to zero, and accelerate
the plasma crossing them to  $V_A $
in the $x $ direction to match the plasma
flowing out of the diffusive region with the
same Alfven $x $ velocity.

	The shocks propagate in the $ y $ direction
upstream into the plasma with velocity
$B_y/\sqrt{4 \pi \rho} $.  Since the plasma
is flowing with the reconnection velocity
$V_y = - V_A $ we have in steady state,
\begin{equation}
\frac{B_y}{\sqrt{4 \pi \rho}} = V_R    ,
\end{equation}
which determines the magnitude of 
$B_y $ the transverse field component.
This $y$ component of the field increases
linearly along $x$  in the diffusive
region from $0$ to this value, and it turns out
that the tension produced by the $ j_z B_y $
force is just enough to accelerate the plasma 
in the layer up to the Alfven speed.

	These results are all given in Petschek's
paper and present a nearly complete, but
qualitative,  physical picture for magnetic
reconnection, encompassing the possibility of 
a diffusive layer much shorter than $L$.
Further, in his theory, $L' $ appears to be
a free parameter.  Petschek then chooses 
$L'$ as short as possible to get the maximum
reconnection velocity.  He determined
 this minimum  to be the lower limit so that
the current in the shocks did not seriously
perturb the incoming magnetic field $B_x = B_0 $.
This limit was roughly 
\begin{equation}
L' > \frac{L}{S} (\ln{S})^2  ,
\end{equation}
so that substituting in equation (7) he
got a very fast limit on the reconnection
velocity
\begin{equation}
V_R < \frac{V_A}{ \ln{S}}   .
\end{equation}
This latter limiting velocity has been generally
quoted in the literature as 
the so-called Petschek reconnection velocity,
and it is this velocity that has been compared
with the Sweet-Parker reconnection velocity
equation (4) in the controversy between the 
two theories.  The disagreement should
more appropriately be between the Sweet-Parker
velocity and equation (7).

	Now, it turns out that $L'$ is not
a free parameter in Petschek's theory.  
There is an additional condition associated
with $B_y$ that Petschek did not include .
  The all important $B_y $ field,
which is needed to support the shocks, is 
embedded in the rapidly moving plasma and is 
swept down stream at the Alfven velocity, $V_A$.
This field which is being swept away so rapidly
must be regenerated at the same rate to
preserve a steady state.

	This regeneration occurs during the merging
process.  The external field is nonuniform,
being strongest near $x=0$, so the lines of
force at the center of the diffusive layer
will move into the diffusive region fastest
there.  However, the nonuniformity of the external
field is on the scale of the length of the layer
$L$ because of the global conditions, so as
$L'$ gets smaller, the nonuniformity over the shorter
length is also smaller, and the regeneration process 
is weaker.  A balance between the nonuniform
merging process that creates the $B_y$ field,
and the down sweeping which destroys it, must be reached
and this leads to a relation between $L'$ and
$B_y$.  Thus, combining this relation with
equations (7) and (8), determines $L'$ uniquely.

Let us estimate this balance
qualitatively.  The equation for $B_y$ will
be shown to be
\begin{equation}
\frac{d B_y}{dt} = \frac{V_R}{L'}\frac{L'^2}{L^2}
B_0  - B_y \frac{V_A}{L'}=0.
\end{equation} 
The second term on the right is the down sweeping
term that destroys $B_y$.  Its form is obvious.

	The first term represents
nonuniform merging, and its form can be derived
as follows: (see Figure 4).

\begin{figure}[t]
\vspace{7.3cm}
\includegraphics{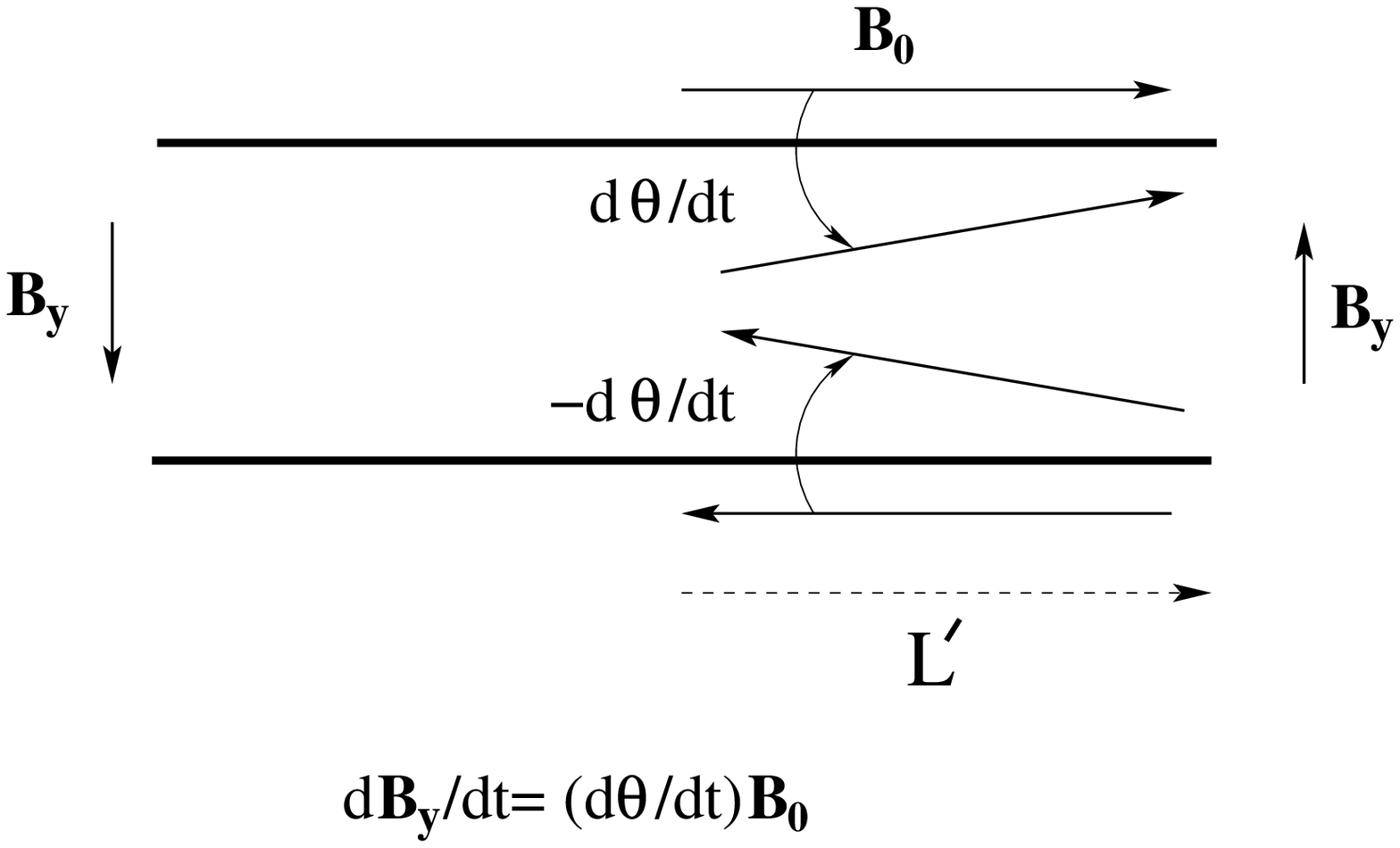}
\caption{ Regeneration of the $ B_y $
field by nonuniform merging. This rotates
the $ B_x $ field into the $ y $ direction.}
\end{figure}

	The external field depends on $x$ 
as
\begin{equation}
B_x =  B_0(1- x^2/L^2)    .
\end{equation}
We assume that each fresh line that is merging 
enters the layer with velocity, $V'_x $ 
proportional to $ B_x(x) $, 
so that the line enters faster at 
$x = 0 $  than at $x = L'$.
Thus, after entering the layer it will
turn at the rate

\begin{equation}
\frac{d \theta}{dt} = \frac{V'_x(0)
-V'_x(L')}{L'} = \frac{\eta c}{4 \pi \delta L'}
[1 -(1-L'^2/L^2)]    ,
\end{equation}
or
\begin{equation}
\frac{d \theta}{dt} =  \frac{\eta c}{4 \pi \delta}
\frac{L'^2}{L^2} \frac{1}{L'}  .
\end{equation}
The turning of a line of strength $B_0 $
at the rate $ d \theta/d t$ produces
a component $B_y $ at the rate
$ (d \theta/dt) B_0 = V_R (L'^2/L^2)B_0/L' $,
which gives the first term in equation(11).

Now, setting $d B_y/dt =0 $, for a steady
state, gives from equation(9),
\begin{equation}
B_y = \frac{V_R}{V_A} (L'^2/L^2) B_0  .
\end{equation}
From equation (8) we get
\begin{equation}
V_R = \frac{V_R}{V_A} ( L'^2/L^2)
\frac{B_0}{\sqrt{4 \pi \rho}} = V_R
( L'^2/L^2) .
\end{equation}
or,
\begin{equation}
L' =L
\end{equation}.

	Thus, $L' $ is no smaller than
$L$, and the Petschek rate equation (7)
reduces to the Sweet-Parker rate
equation (4).

	I believe that this is the
reason that the numerical simulations
always yield the Sweet-Parker rate,
rather than the faster rate implied
by Petschek's formula.

	A more formal derivation of 
equation (11) is given in the appendix.

\section{Anomalous Resistivity in the
Sweet-Parker Model}

	In the absence of a $B_z$
component (that is no guide field), there 
is a strong instability, the lower
hybrid instability, that should be excited,
(Davidson 1975)
This is the case 
if the current density in the layer is
large enough that the difference 
in the electron and ion bulk velocities
$v_{-}$ and $v_{+} $
is greater than the ion acoustic speed.
That is, if the drift velocity
$v_d \equiv v_{+}-v_{-} $ satisfies
$v_d > v_i $ with $v_i $ the ion 
thermal velocity. Note that $j = n_e
(e/c) v_d $.

	There are three 
well-documented instances where
magnetic reconnection is definitely
taking place, i.e. the solar flare,
the magnetosphere-solar wind interface,
and the magnetotail.
	If one examines these  three
cases,  and applies the Sweet-Parker
model to them, one finds in all three
cases that the drift velocity, $v_d$,
is much larger than $v_i $.  One can 
express this as follows: There is
a critical current, $j_c $, and 
critical layer thickness $ \delta_c
= B_0/4 \pi j_c $
such that if $\delta < \delta_c $
and therefore $ j $ is greater than
the critical $j $, then
the lower hybrid mode should be excited.
This lower hybrid instability has
the property, that it can generate
an almost unlimited amount of resistive
friction between the electrons and the
waves.

	Now, let us imagine the two
plasmas with opposite magnetic fields, $B_0$,
approach each other. The pressure, 
between them, $ p= B_0^2/8 \pi $,
 is dissipated at the
rate $V_A/L $ by  expansion due
to flow out the ends, and by force 
balance, must be replenished by
compression due to $d \delta/d t < 0$.
This compression normally continues
until the Sweet-Parker thickness is
reached.
At this time, the plasma pressure
in the layer is replenished by 
Ohmic heating $ \eta j^2 \approx
\eta B_0^2/(4 \pi \delta)^2 $
at the same rate at which it is 
depleted, $ (V_A/L)p= (V_A/L)
(B_0^2/8 \pi) $, by the adiabatic expansion.
Therefore at this time the collapse
$d \delta/d t $ ceases.

	On the other hand, if the critical
thickness, $\delta_c $, is passed
before the Sweet-Parker thickness is
reached, the resistivity rapidly rises
to generate an Ohmic heating large enough to
balance the outflow adiabatic expansion
at this larger distance
$ \delta_c $, and collapse ceases at this larger
distance.

	For these conditions, the layer thickness
 is known, and $V_R $ is 
determined by the mass conservation equation
(2) alone,
\begin{equation}
V_R = \frac{\delta_c}{L} V_A   .
\end{equation}
Thus, reconnection can become much 
 faster than the Sweet-Parker rate 
based on Spitzer resistivity.
Under solar flare conditions, Kulsrud (1998),
 it can
become as much as a factor of a
thousand faster.  The resulting
reconnection time can be reduced to 
a few hours, perhaps an order of
magnitude longer than the
observed energy release time in 
solar flares.

\section{Petschek Reconnection with
Anomalous Resistivity}

	In the second section it was
shown that for a constant resistivity,
Petschek's $L'$ parameter must be
equal to $L$, so that Petschek's reconnection
rate reduces to that of Sweet-Parker.
However, if $\eta$ is anomalous,
enhanced by wave interactions,
it can be very sensitive to the current
density. The original problem with
Petschek reconnection was that the 
external field at $x=L'$ was only
slightly smaller, by a factor of 
$1-L'^2/L^2 $, than its strength
at $x=0$.  But even this slight change in the
resulting current density can lead
to a finite and even large change in the
resistivity $\eta$.  Taking this into
account, one finds that equation (11) 
becomes
\begin{equation}
\frac{d \theta}{d t} =
\frac{\eta_0 c}{4 \pi \delta L'}-
\frac{\eta' c}{4 \pi \delta' L'} 
(1-\frac{L'^2}{L^2})
\approx
\frac{(\eta_0-\eta')c}{4 \pi \delta L'}  ,
\end{equation}
in which we have neglected $L'^2/L^2$
and any slight difference between
$\delta$ and $\delta'$.  $\eta_0$
is the resistivity at $x =0 $, and
$\eta'$ that at $x=L'$.  Solving for
$B_y =( d \theta/dt) B_0 $ as before,
with this different value of $d \theta/dt$,
and using it in Petschek's formula for the
shock velocity, we find that with
 variable resistivity,
\begin{equation}
V_R = \frac{B_y}{\sqrt{4 \pi \rho}}=
\frac{1}{V_A}\frac{B_0}{\sqrt{4 \pi \rho}}
\frac{(\eta_0 -\eta')c}{4 \pi \delta}  .
\end{equation}
Taking $\delta = \delta_c $, 
we have
\begin{equation}
V_R = \frac{c(\eta_0-\eta')}{4 \pi \delta_c}  .
\end{equation}

	Now, to estimate the value
of this revised reconnection
velocity,  we assume that $\eta$ is
linear in $j$ for $j>j_c$,
so that
\begin{equation}
\eta_0 -\eta' =
(j_0-j') \frac{d \eta}{d j} =
\frac{1}{4 \pi}
(\frac{B_0}{\delta}-\frac{B'}{\delta'})\frac{d \eta}{d j}
=\frac{1}{4 \pi} 
\frac{L'^2}{L^2}\frac{B_0}{\delta_c}\frac{d \eta}{d j}
\end{equation}

	Combining this with the mass conservation
relation for the $L'$ layer, 
\begin{equation}
\frac{V_R}{V_A} = \frac{\delta_c}{L'}  ,
\end{equation}
we obtain 
\begin{equation}
\frac{V_R^3}{V_A^3} = \frac{B_0}{V_A L^2} 
\frac{c}{(4 \pi)^2}
\frac{d \eta}{d j}    .
\end{equation}

	This result can be written in a more
familiar way by assigning a maximum value,
$\eta^{*}$ to $\eta$ and assuming that
$\eta = \eta_{spitzer} $ for $j<j_c$,
and $\eta = \eta^{*}$ at $j = 2 j_c$.
Thus, 
\begin{equation}
\frac{d \eta}{d j} = \frac{\eta^{*}}{j_c}  .
\end{equation}
From  $j_c = B_0/4 \pi \delta_c$ equation (24)
reduces to 
\begin{equation}
\frac{V_R}{V_A} = \left( \frac{\delta_c}{L}
 \frac{1}{S^{*}} \right)^{1/3}   ,
\end{equation}
where $S^{*}$ is the modified Lundqvist number
based on $\eta^{*} $
\begin{equation}
S^{*} = \frac{V_A L}{\eta^{*} c/4 \pi}
\end{equation}

	Numerically, $\eta^{*} $ comes from
an electron wave collision rate equal to
the electron plasma frequency $\omega_{pe}$.
Under typical solar flare conditions,
Kulsrud (1998),
$\eta^{*} \approx 10^6 \eta_{Spitzer}$,
and
\begin{equation}
\frac{V_R}{V_A} \approx 10^{-4}
\end{equation}
One can carry out similar estimates
for the magnetosphere-solar wind
interface and one finds from
 equation (26) that
\begin{equation}
\frac{V_R}{V_A} \approx 10^{0}
\end{equation}

\section{Conclusions}

	We have shown or stated that:

(1). In general  reconnection situations,
$L$ and $B_x$ are determined globally,
while $\delta$ and $V_R$ are determined
locally.

(2). For constant resistivity, the length
of Petschek's diffusive layer is not a
free parameter, but is determined by
the condition that $B_y$ be regenerated
at the same rate as it is being dissipated
by down stream flow.

(3).Constant resistivity  gives $L'=L$,
which makes Petschek's reconnection rate
equal to that of Sweet and Parker.

(4). If the Sweet-Parker thickness $\delta
= L/\sqrt{S} $ is thinner than the 
critical thickness $\delta_c$ at which
anomalous resistivity sets in, then
the Sweet Parker reconnection rate becomes
\begin{equation}
V_R = \frac{\delta_c}{L}V_A   ,
\end{equation}
a rate that can be very much faster than
their reconnection rate based on Spitzer
resistivity.

(5). In the case of anomalous resistivity
the regeneration rate of $B_y$ in Petschek's
theory is much larger, and the Petschek's
rate becomes faster even than the Sweet-Parker
rate with enhanced resistivity.  It is given by
\begin{equation}
\frac{V_R}{V_A} = 
\left( \frac{\delta_c}{L}\frac{1}{S^{*}}\right)^{1/3}
\end{equation}
where $S^{*} = L V_A/(\eta^{*} c/4 \pi) $
is the Lundqvist number based on the maximum
possible resistivity $\eta^{*}$.
Note that it has a cube root dependence
on this maximum resistivity, rather than
a logarithmic dependence on the Spitzer
resistivity which is  the often quoted
expression for Petschek reconnection.
In spite of this,  in many cases there is
not a large numerical difference in the two results.
Formula (31) gives an equally fast reconnection rate,
and is more in tune with the true physical
processes.

(6). A test for whether the anomalous
resistivity rate equation (18) or (31), rather than
 the classical Sweet-Parker rate,
 equation (7), is applicable
is:  First, compute the Sweet-Parker
thickness $\delta_{SP} $, of the reconnection layer 
$\delta_{SP} = L/\sqrt{S} $, and compare it with
the critical thickness $\delta_c= B_0/(4 \pi n_e v_i/c)
$.  If $\delta_{SP} < \delta_c $, then use the anomalous
equation (31) for Petschek reconnection, or the
anomalous Sweet-Parker equation (18), whichever
is faster.

(7). In nearly all cases on the galactic scale,
$\delta_{SP} $ is larger than or at least comparable
to $\delta_c $, so the Sweet-Parker result
gives the correct order of magnitude for the 
reconnection rate.  This is almost always too
slow to be of interest, so  one concludes that
reconnection on the galactic scale is hardly 
ever really important.

\section{Acknowledgment}
I gratefully acknowledge many useful discussions
with my colleagues, Dmitri Uzdensky,
 Masaaki Yamada and Hantao Ji.
I am also very grateful for much help
from Leonid Malyshkin in preparing the
manuscript for publication.

\appendix

\setcounter{equation}{0}
\renewcommand{\theequation}{A\arabic{equation}}

\section*{Appendix}

	In this appendix we justify the
intuitively described  equation (11) for
the evolution $B_y$, by a more precise derivation.
In the intuitive derivation, it was assumed
that the lines flowed into the reconnection
only by resistive merging, and the
effect of plasma flow on them was ignored.
Also, the merging velocity was taken proportional
to the external field, through its effect
on the current density.  Further, the thickness
of the reconnection layer was assumed constant
between $x=0$ and $x=L'$.

	Consider an $x$ dependent thickness
$\delta(x)$, as shown in figure 5, taken large
enough that at $ y = \delta(x), 
B_x \approx B(x)=B_0(1-x^2/L^2) $,
the value for the external field.  Also take
$\delta(x)$  to follow a line of force.
We have Ohm's law in the layer,
\begin{equation}
-c E_z=- V_y B_x +V_x B_y -\eta c j_z    .
\end{equation}
In a steady state $E_z  $ is a constant.

\begin{figure}[t]
\vspace{7.0cm}
\includegraphics{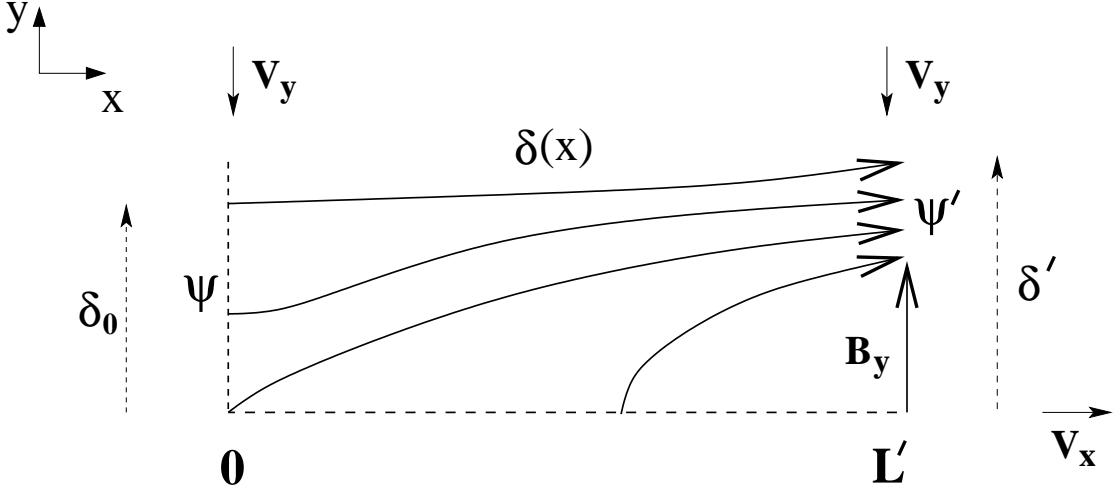}
\caption{Sketch of the Petschek
diffusion layer for
 the $ B_y $ field derivation.
At $ x = L' , B_x $ is zero nearly
up to $ y = \delta' $. $ \delta(x) $
is a line of force.}
\end{figure}

	Integrate $- c E_z $ at $x = 0$,
from $y=0$ to $y= \delta_0 = \delta(0) $.
See figure 5. Since $B_y = 0$ at $x=0$, we have
\begin{equation}
c \int_0^{\delta_0} -E_z dy =
\int_0^{\delta_0} -v_y B_x dy +\eta c \int_0^{\delta_0}
-j_z dy   ,
\end{equation}
or
\begin{equation}
c \int_0^{\delta_0} -E_z dy =
\int_0^{\delta_0} -v_y B_x dy +
\frac{\eta c}{4 \pi} \int_0^{\delta_0}
 \frac{\partial B_x}{\partial y} dy   ,
\end{equation}
or
\begin{equation}
c \int_0^{\delta_0} -E_z dy =
\int_0^{\delta_0} -v_y B_x dy +
\frac{\eta c}{4 \pi}  B_0    .
\end{equation}
Note that $-E_z ,-V_y$, and $-j_z$ are positive.

	Correspondingly,  integrate $-E_z $ at
$x=L'$,  from $y=0$ to $y=\delta'=\delta(L')$,
\begin{equation}
c \int_0^{\delta'} -E_z dy =
\int_0^{\delta'} -v_y B_x dy +\int_0^{\delta'} v_x B_y dy +
\frac{\eta c}{4 \pi} B_0(1-L'^2/L^2)   .
\end{equation}

	At large enough $\delta$, we have
\begin{equation}
-V_y(0,\delta_0) =V_R, 
\end{equation}
and
\begin{equation}
-V_y(L',\delta') =\frac{V_R B_0}{B_0 (1-L'^2/L^2)}, 
\end{equation}
since $y=\delta(x) $ is in the infinitely
conducting region, and
\begin{equation}
-c E_z =-V_y(0,\delta_0) B_x(0,\delta_0)
= -V_y(L',\delta') B_x(L',\delta')  .
\end{equation}

But, because $E_z$ is constant for all $x$ and $ y$,
the left hand side of equation (A4) and (A5) are equal
to $- c E_z \delta_0 $ and $- c E_z \delta' $,
respectively.  Therefore, subtracting the right
hand side of equation (A5) from $\delta'/\delta_0$
times that of  equation (A4),  we get
\begin{equation}
\frac{\delta'}{\delta_0}
\int_0^{\delta_0} -v_y B_x dy -\int_0^{\delta'} -v_y B_x dy +
\frac{\eta c}{4 \pi} 
\left[\frac{\delta'}{\delta_0}B_0-B_0(1-\frac{L'^2}{L^2})\right]
=\int_0^{\delta'} -v_y B_x dy .
\end{equation}

	The third term  on the left hand side,
the $\eta$ term, is the non uniform merging term
quoted in the text.  In fact, we can write it as
$\alpha (\eta c/4 \pi) (B_0/\delta_0)(L'^2/L^2) $
where $\alpha$ is a constant of order unity
and equal to $1 + (1/2) d^2 \delta/d x^2$,
which includes the change of $\delta(x)$ with 
respect to $x$. The right hand side 
represents the down sweeping term which destroys
$B_y$.  

We now argue that the sum of the
 first two terms on the left
hand side is negative, so that the expression in
the text for $ B_y$ should be an inequality rather
than an equality.  Since this is partially compensated
by the fact that the regeneration term is slightly
larger, by the factor of $\alpha >1 $, we 
work with equality in the main text.

	First, the mean value of $-V_y$ in the first
term, weighted by $B_x$ is about 2/3 times
$ -V_y(\delta_0) = V_R $
because both $-V_y$ and $B_x$ are linear in $y$.
Thus,
\begin{equation}
\frac{\delta'}{\delta_0}
\int_0^{\delta_0} -v_y B_x dy \approx 
\frac{\delta'}{\delta_0} \frac{2}{3} V_R \psi
\end{equation}
where 
\begin{equation}
\psi = \int_0^{\delta_0}  B_x dy   , 
\end{equation}
is the $B_x$ flux up to $\delta_0$.
On the other hand,
we see from figure 5 that the entire $B_x$ flux
at $y=L'$ is near $\delta'$,  so  here
the $B_x$ averaged value of $V_y$ is
$V_R/(1-L'^2/L^2) \approx V_R$.
Thus the sum of the first two terms in equation (A9)
is essentially,
\begin{equation}
\frac{2}{3} \frac{\delta'}{\delta_0} V_R \psi
-\frac{V_R}{1-L'^2/L^2} \psi'
\end{equation}
where $\psi'$ is the $B_x$ flux at $y=L'$.
Further, we also see, from Figure 5, that $\psi'$
is greater than $\psi$ because $\psi'$
contains the $\psi$ flux (because 
$\delta(x) $ is a line of force)
plus the flux through $y=0$, between 
$x=0$ and $x=L'$.  Therefore, if
$\delta'/\delta < 3/2 $, a  plausible
assumption, since $L' \ll L$, then
the sum in equation (A12) is negative.
In addition, the factors $\psi'/\psi >1$
and $1/(1-L'^2/L^2) >1 $ reinforce the
conclusion that the sum is negative.

	Now, dropping the first two terms
in equation (A9) whose sum has been shown
to be negative, we obtain the inequality,
\begin{equation}
\alpha \frac{\eta c}{4 \pi} B_0 \frac{L'^2}{L^2} >
\int_0^{\delta_0} -v_x B_y dy \approx
V_A \delta' <B_y>
\end{equation}

	Comparing this with equation (11) of
the text we see that, if $\alpha $
were one,   equation (15) would yield
 an overestimate of $B_y$.  Since 
$\alpha $ is certainly of order one,
it is safe to take equation (15) 
as an estimate of $B_y$.

	Note, that equation (A9)
is approximately the rate of change
of $L' \delta B_y $ since
\begin{equation}
\int_0^{L'} dx \int_0^{\delta(x)} dy \frac{\partial B_y}
{\partial t} =
\int \int dx dy \frac{\partial E_z}{\partial x} =
\delta(x) E_z|_0^{L'} ,
\end{equation}
and the latter expression can easily be identified
with the difference of the two integrals given
in equations (A4) and (A5) with the extra
$ \delta'/\delta_0$ factor.

\end{document}